\documentclass[a4paper,11pt]{article}
\usepackage{graphicx}
\usepackage{graphics}
\title{Angular anisotropy of the fusion-fission and quasifission
 fragments} 
\author{\bf A.K.~Nasirov$^{1,2}$,
 A.I.~Muminov$^{2}$, R.K.~Utamuratov$^{2}$,
G.~Fazio$^3$,\\
{\bf G.~Giardina$^3$, F.~Hanappe$^4$, 
G.~Mandaglio$^3$, M.~Manganaro$^3$, and W.~Scheid$^5$}\\
$^1$Flerov Laboratory of Nuclear Reactions, JINR, 
Dubna, Russia\\
$^2$Heavy Ion Physics Department, Institute of
Nuclear Physics, Tashkent, Uzbekistan\\
$^3$ INFN, Sezione
di Catania, \\and Dipartimento di Fisica dell'Universit\`a di
Messina, Messina, Italy\\
$^{4}$Universit\'e Libre de Bruxelles,
 Bruxelles, Belgium\\
$^{5}$Institut f\"ur Theoretische Physik
der Justus-Liebig-Universit\"at, Giessen, Germany} 
\begin{document}
\maketitle
\begin{abstract}The anisotropy in the angular distribution of the
fusion-fission and quasifission fragments  for the
 $^{16}$O+$^{238}$U, $^{19}$F+$^{208}$Pb and $^{32}$S+$^{208}$Pb reactions
  is studied
by analyzing the angular momentum distributions of the dinuclear
system and compound nucleus which are formed after capture and
complete fusion, respectively. The orientation angles of axial
symmetry axes of colliding nuclei to the beam direction are taken
into account for the calculation of the variance of the projection
of the total spin onto the fission axis.
 It is shown that the deviation of the experimental
angular anisotropy from the statistical model picture   is
connected  with the contribution of the quasifission fragments
which is dominant in the $^{32}$S+$^{208}$Pb reaction. Enhancement
of anisotropy at low energies in the  $^{16}$O+$^{238}$U reaction
is connected with quasifission of the dinuclear system having low
temperature and effective moment of inertia.
\end{abstract}
\par
PACS:      25.70.Jj-Fusion and fusion-fission reactions;
      25.70.Lm-Strongly damped collisions; 
      25.85.Ge-Charged-particle-induced fission,  capture, fusion, 
      quasifission,  angular anisotropy of fragments.
\par
%
\newpage
\section{Introduction}
\label{intro} The study of the  mechanism of the  fusion-fission
process in the reactions with massive nuclei is of  interest  of
both experimentalists and theorists to obtain a favorable way of
the synthesis of superheavy elements or exotic nuclei far from the
stability line. The last experiments on the synthesis of
superheavy elements $Z$=114, 115, 116 and 118 were successful at
beam energies corresponding to 35-40 MeV excitation energies of
the compound nucleus which is enough higher than the Bass barrier.
The deformed actinide nuclei were used as  targets at synthesis of
new superheavy  nuclei in the $^{48}$Ca + $^{238}$U, $^{244}$Pu,
$^{248}$Cm reactions \cite{Ogan}. It means that the orientation
angle of the symmetry axis of target-nucleus relative to the beam
direction affects on the fusion-fission mechanism. The cross
sections of events corresponding to the synthesis of
 superheavy elements are not higher than few picobarns \cite{Ogan} and the
width of evaporation residues excitation function is very narrow.
At the same time the measured cross sections of the fission
fragments are several tens of millibarn \cite{Itkis} and  the
excitation function of fission fragments yields is very wide. It
means that only a very small part of collisions in the narrow
range of the beam energy leads to the formation of the evaporation
residues considered as superheavy elements. The problem is to
establish this small range of beam energy as the optimal condition
for the  synthesis of superheavies.
 The main reason leading to the small values of the
evaporation residue cross-sections seems to be connected with the
small survival probability $W_{sur}$  of the heated compound
nucleus against fission by evaporating neutrons. It is well known
that the $W_{sur}$ decreases  by an increase  of the excitation
energy $E^*_{CN}$ and angular momentum $\ell_{CN}$ of the compound
nucleus  \cite{FazioEPJA22}.

 But the formation of the compound nucleus in
reactions with massive nuclei has a hindrance: not all of the
dinuclear systems formed at capture of the projectile by the
target-nucleus can be transformed into compound nuclei. We should
stress that the estimation of the formation probability is
difficult by both experimental and theoretical methods. The
determination of the fusion probability from the experimental data
is ambiguous  due to the difficulties to identify pure fission
fragments of the splitting compound nucleus from the fragments
which are formed in other processes of heavy-ion collisions like
fast-fission, quasifission and deep inelastic collisions. By the
way, restoration of its value from the cross-sections of
evaporation residues is model-dependent. As a result there is a
field for speculations which can be clarified indirectly  by the
analysis of the physical results connected with the formation of
the compound nucleus. The angular distribution of reaction
fragments is one of the  informative quantity allowing us to study
the fusion-fission mechanism of heavy-ion collisions.

The goal of the present paper is to show the ability of the method
based on the dinuclear system (DNS) concept to calculate the
angular momentum  distribution for the fusion-fission and
quasifission fragments by analyzing the anisotropy of the angular
distribution of both fusion-fission and quasifission fragments in
reactions with deformed and nearly spherical target  nuclei.
  The partial capture, fusion and quasifission excitation
functions, as well as the corresponding mean square values of
angular momentum calculated  in this work were used to determine
the anisotropy $A$ of the  fragment angular distribution by
formula (\ref{anisA}) as a function of  the spin distribution of
the fissioning  systems: compound nucleus and dinuclear system.
The results were compared with the experimental data for the
observed anisotropy $A$ of the angular distributions of fragments
of the $^{16}$O+$^{238}$U, $^{19}$F+$^{208}$Pb and
$^{32}$S+$^{208}$Pb reactions.

 The paper is organized in the following way.
 In sect. \ref{about}, we discuss the possibility of the use of
the anisotropy of the angular distribution of fragments to
establish their origination. In sect. \ref{Msv} we present  the
method to calculate the orbital angular momentum distribution,
mean square values $<\ell\,^2>$,   and   anisotropy $A$  of the
angular distribution of the fission and quasifission fragments. A
short presentation about how we calculate capture, fusion, and
quasifission excitation functions is given in sect. \ref{method}.
 The results of the anisotropy of the fragment angular
distribution contributed by the fission and quasifission are
calculated and discussed in sect. \ref{discuss}. Conclusions are
given in sect. \ref{concl}.

\section{About the interpretation of the anisotropy of
angular distribution of reaction fragments} \label{about}

Generally, authors analyzing the experimental data determine the
fission cross sections   by fitting of the measured  angular
distributions of the fission fragments. To generate the angular
momentum distributions required to predict the shape of the
fission angular distributions,  the approximate fusion cross
sections is used (see, for example, ref. \cite{Hinde}). In this
paper, the ''restored" fission angular distributions were used
again and fitted to the measured data for the $^{19}$F+$^{208}$Pb
reactions. From the final fits the anisotropies $A$ and the
fusion-fission cross sections as well as the value of $K_0^2$ at
each bombarding energy were determined. In such an analysis, it
was implicitly assumed that $K_0^2$ is independent {\bf on} the total
spin $J$. The main conclusion of the authors  was that the
anisotropy at the highest bombarding energies can only be
reproduced by assuming that the fission barrier does no longer
control the fission process when its height is less than the
nuclear temperature. The role of  the quasifission process was not
discussed in \cite{Hinde}. In ref.\cite{Hinde53}, the fission
fragment anisotropies and mass distributions were measured over a
wide range of angles for the $^{16}$O+$^{238}$U reaction. The
authors concluded that a systematic deviation of the measured
fission fragment anisotropies from the transition state model
predictions  confirms the validity of  the correlating anomalously
large anisotropies with the presence of quasifission. In
refs.\cite{BackPRL,Back}, the sensitivity of the features of
fission-fragment angular distributions to nonequilibrium processes
such as quasifission was shown by a quantitative analysis of the
angular distributions of near-symmetric masses produced in the
$^{32}$S+$^{208}$Pb reaction.

The study of correlations between mass and angular distribution of
fragments of full momentum transfer reactions allows us to
separate the pure fission fragments of the compound nucleus with a
compact shape \cite{Shen}. The mass and angular momentum
distributions of the reaction fragments are determined by the
dynamics of collision. The mass distribution strongly depends on
the potential energy surface. Quasifission produces fragments
alike the fission fragments confusing the estimation of the fusion
cross section. In the quasifission process, the  compound nucleus
stage is not reached. The complete kinetic energy relaxation
(capture stage) is a main characteristics of the quasifission
reactions. It means that quasifission takes place only after the
capture of the projectile by the target-nucleus. The mass
equilibrium can be reached or not in dependence on the masses and
mass asymmetry of the reactants \cite{Back}, as well as on  the
dynamics of collision.

The symmetric mass distributions at all angles and symmetric
angular distributions relative to $\theta_{c.m.}=90^{\circ}$ are
characteristic features for the fission decay of the completely
fused system. The angular distributions of fission fragments are
often characterized by  the ratio of the yield at $180^{\circ}$
(or $0^{\circ}$) to the one at $90^{\circ}$, {\it i.e.},
$A=W(180^{\circ})/W(90^{\circ})$. The angular distribution of the
fission products is described in the framework of the standard
statistical model (SSM) usually making use of the fact that the
fission saddle-point configuration can be treated as a transition
state between the compound system in its quasi-equilibrium state
and the two separated fission fragments \cite{Vanden}. This model
is used under the assumption that the final direction of fragments
is given by the orientation of the nuclear symmetry axis as the
nucleus passes over the fission saddle-point. This assumption may
not be justified for the heaviest systems for which the saddle-
and scission-point configurations have very different shapes.
Consequently, in such case, the SSM may not describe properly the
angular anisotropy of pure fission fragments in reactions with
massive nuclei. Certainly, the discrepancy between theoretical and
experimental estimations of the angular anisotropy is caused by an
imperfection of the SSM and the  experimental difficulties of
separating the pure fission fragments.

The models which reproduce the fusion-fission excitation functions
fail to account for the fission-fragment angular-distribution
data. The reasons of the failure could be connected with the
effects of the entrance channel (presence of quasifission)  or the
fission exit channel ($K$-equili- brium is not reached, where $K$ is
the projection of the total spin of the nucleus on its axial
symmetry axis) \cite{Ramam85,Vopkap95}.
  Certainly  effects of  both of above-mentioned
phenomena should be analyzed with the increase of the anisotropy
$A$ in the angular distribution of reaction fragments. Vopkapic
and Ivanisevic in ref. \cite{Vopkap95} suggested that at
sub-barrier energies, fusion of projectile occurs only when the
prolate deformed target is oriented in the beam direction,
producing a narrow initial $K$ distribution peaked around $K=0$.
The $K$ equilibration time was also assumed to be not too short
compared to the fission time. Using of a time dependent and narrow
$K$ distribution compared to  predictions of the statistical
saddle-point model could be envisaged and the fragment angular
anisotropy could be explained.

The deviation of the experimental angular anisotropy from the
statistical model picture at energies above the interaction
barrier was discussed by authors of  ref. \cite{Hinde}. The reason
of the failure of the statistical model to reproduce the measured
data was considered doubtful for the application of the SSM at
high energies when it is no longer valid if the fission barrier
could be passed at the first attempt. This deviation occurs at
large values of angular momenta where the fission barrier height
is less than the saddle-point temperature ($T \approx 1.6$ MeV).
It means that all the properties of the fission process should be
determined completely by the dynamics of the motion over the
potential energy surface (PES).

The measured fission data corresponding to large values of the
fragment angular anisotropy $A$ can include a contribution of
quasifission fragments leading to higher anisotropy than the ones
predicted by standard  statistical models \cite{Hinde,Back} since
at quasifission the dinuclear system  never becomes as compact to
be the compound nucleus, and also the $K$ equilibration  is
probably not attained. The experimental data (see,  for example,
ref. \cite{Toke}) confirm events with characteristic features,
particularly in association with projectiles heavier than
$^{24}$Mg.

So, there are two main points of view to interpret the
experimentally observed angular anisotropy $A$: 1) authors  of
refs. \cite{Hinde,Back} and we, in the present paper, explain it
with the contribution of the quasifission process competing  with
the formation of the compound nucleus; 2) authors of ref.
\cite{Zhang} observe a strict evidence of the anisotropy $A$,
explaining such an anomaly by a new version of the preequilibrium
fission model \cite{Zhang2}.

 Calculations within the SSM assume the availability of
 a realistic spin distribution of the fissioning system.
 In turn the calculation of the spin
distribution of the compound nucleus formed in the heavy-ion
induced reactions is a complicated task. The extraction of a
realistic spin distribution from the measured angular distribution
of reaction fragments may be ambiguous due to a sufficient
contribution of quasifission.

 We consider the role of the entrance channel  in the observed angular
anisotropy.  The mean square values $<\ell^2>$ versus $E_{c.m.}$
is determined. We compare our results of $A$ with the available
estimations extracted from the experimental data
\cite{Hinde,Back}.

\section{Calculation of the anisotropy  and mean square values  $< \ell^2 >$
of the angular distribution} \label{Msv}

The angular distribution of splitting fragments of the rotating
system is determined by its angular momentum distribution, namely,
by the projection, $K$, of the total spin vector, $J$, onto the
center axis of the separated fission fragments and by the moment
 of inertia of the fissioning system. The total spin has
no component along the beam axis ($M=0$), if the fissioning system
is formed at capture (full momentum transfer) of a spinless
projectile by a spinless target.We calculate the anisotropy $A$
using our results  of the angular momentum distributions
$<\ell\,^2>$ for the complete fusion and quasifission, and
$\mathcal{J}_{eff}$ for compound nucleus is found by the rotating
finite range model (RFRM) by Sierk \cite{SierkPRC33} and for the
DNS is determined by our model taking into account different
mutual orientations of symmetry axes of interacting nuclei.
 Then we can use the  expression
for the approximated anisotropy of the fission fragment angular
distribution suggested by Halpern and Strutinski in
ref.\cite{Halpern} and Griffin in ref.\cite{Griffin}:
\begin{equation}
\label{anisA}
 A \approx 1+\frac{<\ell^2>_{fus}\hbar^2}{4\,<\mathcal{J}_{eff}T_{sad}>},
\end{equation}
where
\begin{equation}
\label{Jeff}
 \frac{1}{\mathcal{J}_{eff}}= \frac{1}{\mathcal{J}_{\|}} - 
 \frac{1}{\mathcal{J}_{\bot}}
\end{equation}
 is the effective moment of inertia on
the saddle point for the  compound nucleus; $\mathcal{J}_{\|}$ and
$\mathcal{J}_{\bot}$ are moments of inertia around the symmetry
axis and a perpendicular axis, respectively. Their values are
determined in the framework of the RFRM by Sierk
\cite{SierkPRC33}.  $\mathcal{J}_{eff}$ and $T_{sad}$ are
functions of $<\ell>$ and  their values for the  given beam energy
and orbital angular momentum  are found by averaging
$<\mathcal{J}_{eff} T_{sad}>$ by the partial fusion cross
sections, similar as in formula (\ref{uno}).

The mean square values of the orbital angular momentum for the
fusion-fission $< \ell\,^2 >_{fus}$ and  quasifission
$<\ell\,^2>_{qfiss}$ processes are calculated by using  the
partial  cross sections of fusion,  $\sigma_{fus}^{(\ell)}$, and
quasifission, $\sigma_{qfiss}^{(\ell)}$, respectively. The
above-mentioned mean square values are found by averaging over all
orientation angles of  the symmetry axis of deformed nuclei
\cite{Nasirov}:
\begin{eqnarray}\label{uno}
{<\ell\,^2 (E)>}_{(i)}&=& \frac{\sum_{\ell=0}^{\ell=\ell_d}
\ell\,^2 <\sigma^{(\ell)}_{(i)}>_{\alpha_{P},\alpha_{T}}\,(E)}
{\sum_{\ell=0}^{\ell=\ell_d}
<\sigma^{(\ell)}_{(i)}>_{\alpha_{P},\alpha_{T}}(E)}
\end{eqnarray}
with
\begin{eqnarray}\label{sigmal}
&&<\sigma^{(\ell)}_{(i)}>_{\{\alpha_{P},\alpha_{T}\}}(E)=
\int_0^{\pi/2}\sin\alpha_P\int_0^{\pi/2}
\sin\alpha_T  \nonumber \\
&&\times \sigma_{(i)}^{(\ell)}(E; \alpha_P,\alpha_T) d\alpha_Td\alpha_P,
\end{eqnarray}
where  $i=fus$ or $qfiss$, and $\alpha_P$ and $\alpha_T$ are the
orientation angles of the axial symmetry axes of the projectile
and target nuclei, respectively.

 The effective temperature $T_{sad}$ at the
saddle point is related to the excitation energy by the
expression:
\begin{equation}\label{Tsad}
 T_{sad}=\left[\frac{E_{c.m.}+Q_{gg}-B_f(\ell)-E_n}{A_{CN}/8}\right]^{1/2},
 \end{equation}
where $Q_{gg}$ and $B_f(\ell)$ are the reaction $Q_{gg}$-value for
the ground states of nuclei and the fission barrier height,
respectively. The $B_f(\ell)$ is calculated in terms of the RFRM
by Sierk \cite{SierkPRC33}.  For the given $E_{c.m.}$ we calculate
$<\ell>$ and its value is used to find $B_f(\ell)$. $A_{CN}$ is
the mass number of the composite system and $E_n$ the energy
carried away by the pre-saddle fission neutrons.  The last was not
analyzed  in this work.
 An important physical quantity in formula (\ref{anisA})
is the variance $K_0^2$ of the Gaussian distribution of the $K$
projection:
\begin{equation}\label{formK02}
 K_0^2=\frac{\left < \mathcal{J}_{eff} T_{sad}\right >}{\hbar^2}.
 \end{equation}
Often $K_0$ is used to fit the angular distribution of fission
fragments (see ref.\cite{Back}).

For the estimation of the anisotropy of quasifission fragments  we
calculate $\mathcal{J}_{eff}$  for the dinuclear system taking
into account the possibility of different orientation angles of
its constituent nuclei (see Appendix A). The excitation energy of
the dinuclear system is found as a sum of the difference between
the beam energy and the minimum of the potential well of the
interaction potential and the $Q_{gg}$-value corresponding to a
change of the excitation energy of the dinuclear system from the
projectile-target configuration to the quasifission fragments.
Assuming that after capture the mutual orientations of the DNS
nuclei  do not change much, we calculate $\mathcal{J}_{eff}$ for
collisions of the projectile and target with different
time-independent orientations of their symmetry axes. Under this
assumption, and using our calculated mean square values $<\ell\,^2>$
for quasifission we determine the angular anisotropy $A$ of the
quasifission fragments.  The effective  value $\mathcal{J}_{eff}$
of the moment of inertia of the
 dinuclear system  ($\mathcal{J}^{DNS}$) is found by averaging
 on all the $\ell$ values  and orientations
  $(\alpha_P , \alpha_T)$ with the partial capture cross sections,
  for a given collision energy:
\begin{eqnarray}\label{sei}
\mathcal{J}_{eff}^{(DNS)}&=&
\sum_{\alpha_P,\alpha_T}\sum_{\ell} \mathcal{J}^{(DNS)}
(\ell,\alpha_P,\alpha_T)\sigma_{capt}^{(\ell)}(\alpha_P,\alpha_T)/
\nonumber\\
&&\sum_{\alpha_P,\alpha_T}\sum_{\ell} 
\sigma_{capt}^{(\ell)}(\alpha_P,\alpha_T),
\end{eqnarray}
where
$\sigma_{capt}^{(\ell)}=\sigma_{fus}^{(\ell)}+\sigma_{qfiss}^{(\ell)}$.
The  values of  $< \ell\,^2 >$  for  the fragments of  quasifission
are higher than the ones of the compound nuclei \cite{Nasirov}.
This kind of fission-like decay produces a high anisotropy in the
angular distributions due to the large angular momentum of DNS
\cite{Murakami}, because the partial cross section of quasifission
increases by increasing of $\ell$ \cite{Nasirov,Giardina}. The
reason is that the hindrance for the transformation of the
dinuclear system into compound nucleus
 increases due to an increase of the intrinsic  fusion barrier
 $B^*_{fus}$ with  the  orbital angular momentum $\ell$. At the
 same time quasifission barrier $B_{qf}$ decreases by increasing of $\ell$
  \cite{Giardina,FazioPRC72}.
 We determine  these  barriers of the DNS model in sect.  \ref{fusion}.

 The dissipation of the initial
orbital angular momentum $\ell_0$ of collision  during the capture
process  and the maximum value $\ell_d$ of the  partial waves
leading to capture  are calculated by the solution  of the
corresponding equation of motion. The results show that such a
dissipation is considerable and the value of angular momentum
after dissipation $\ell_f$ is about 25--30\% lower than the
initial value $\ell_0$. This value is found by the solution of the
equations of motion for the orbital angular momentum and radial
motion of nuclei. Details of these calculations can be found in
refs. \cite{Nasirov,GiardEPJ8,SymNas}. The possibility to
calculate the spin distribution of the compound nucleus (its
angular momentum distribution) is the advantage of the used method
based on the dinuclear system concept \cite{Volkov}. In the next
sect. \ref{capture} we present shortly the basic points of the
model.

It should be stressed that, in the case of collisions of
 deformed nuclei, the orientation angles $(\alpha_{P,T})$
 of the symmetry axes   to the beam direction play an important role at the
capture and complete fusion stages. The importance  of the
orientation angles of the symmetry axes of the  reacting nuclei
was analyzed in ref.\cite{Nasirov}. The final results of the
capture and complete fusion are obtained by averaging the
contributions calculated for  different orientation angles of the
symmetry axes of the reacting nuclei with formula (\ref{sigmal}).

\section{Capture, fusion and quasifission cross sections}
\label{method}

\begin{figure*}[h,t,b]
\vspace{0.5cm}
\hspace{-0.7cm}
\resizebox{1.05\textwidth}{!}{\includegraphics{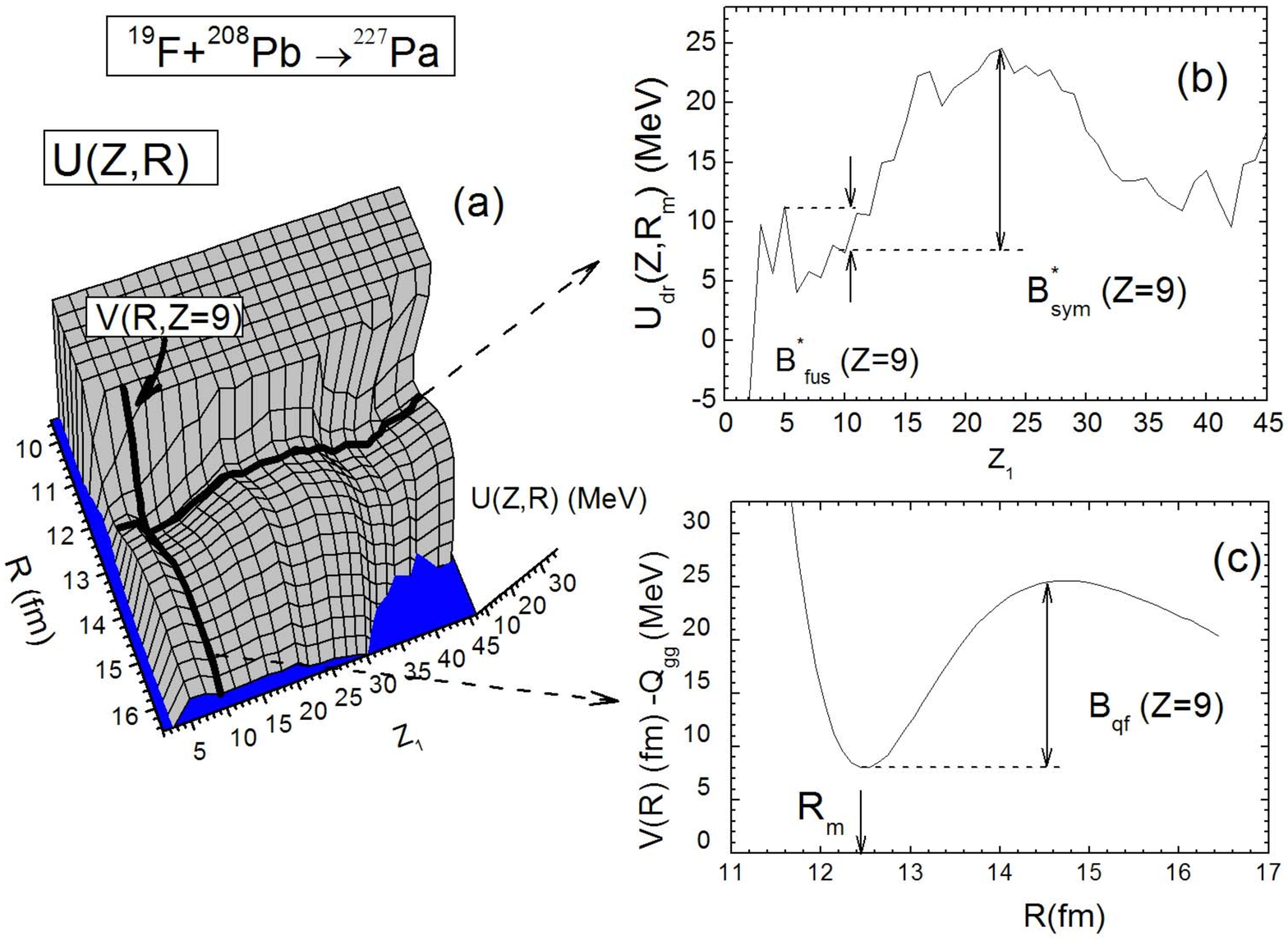}}
\vspace{-0.2cm} \caption{\label{figDNS}  (a) Potential energy
surface for the reactions leading to $^{227}$Pa as a function of
the charge asymmetry of the dinuclear system fragments and the
relative distance between their centers;  (b) driving potential
for the reactions leading to $^{227}$Pa as a function of the
charge asymmetry of the dinuclear system fragments: the intrinsic
fusion barrier $B^*_{fus}$ is shown as the  difference between the
maximum value of the driving potential to the way of complete
fusion and its value corresponding to the considered charge
asymmetry
 of the entrance channel;  the barrier to the mass symmetric
configuration  $B^*_{sym}$ is shown as the difference between the
maximum value of the driving potential to the way of symmetric
masses and its value corresponding to the considered charge
asymmetry  of the entrance channel. (c) The nucleus-nucleus
interaction potential $V(R)$ for the $^{19}$F+$^{208}$Pb system:
the quasifission barrier $B_{qf}$ as the a depth of the potential
well.
} 
\end{figure*}

At the early stage of the reaction with massive nuclei, the
complete fusion of colliding nuclei has a very strong competition
with the quasifission process which decreases the probability of
the compound nucleus formation. The fusion and  quasifission  are
considered  as a two stage process
\cite{Nasirov,Giardina,GiardEPJ8}: (i) the formation of a
dinuclear system as the result of the capture of the
projectile-nucleus by the target-nucleus;  (ii) the transition of
the dinuclear system into the compound nucleus (complete fusion)
as a special  channel of  its evolution. The other alternative way
of the dinuclear system evolution is quasifission. The
quasifission  is the decay of the dinuclear system without
formation of the compound nucleus. Both processes can produce
fragments  with similar characteristics as total kinetic energy
and  mass distributions. The ratio of yields from both channels
depends on the structure of the PES (see, for example, fig.
\ref{figDNS}a) which is different for the different total mass and
charge numbers. In fig. \ref{figDNS}, we presented PES calculated
for reactions leading to $^{227}$Pa. The probability of
quasifission is determined by the relief of PES of the dinuclear
system calculated as a function of the relative distance and mass
asymmetry.  In fig. \ref{figDNS}b, the curve connecting minimums
of the valley on the PES is the driving potential as a function of
the  charge asymmetry of the DNS fragments. The cut of PES for the
given charge number is the nucleus-nucleus interaction potential
$V(R)$. The curve in fig. \ref{figDNS}c was calculated
 for the $^{19}$F+$^{208}$Pb reaction. The size of the potential
 well is determined  by the orbital angular momentum  leading to capture
 and the depth of the potential well is the quasifission barrier
 $B_{qf}$ for the given charge asymmetry. For the interacting
 deformed  nuclei PES depends on the orientation angles
 of the  symmetry axes (see formula \ref{PES}).

\subsection{Capture}
\label{capture}

The partial capture cross section is determined by the capture
probability ${\cal P}^{\ell}_{cap}(E)$ which means that the
colliding nuclei are trapped into the well of the nucleus-nucleus
potential after dissipation of a  part of the initial kinetic
energy and orbital angular momentum:
 \begin{equation}
 \label{parcap}
\sigma^{\ell}_{cap}(E,\ell;\alpha_1,\alpha_2)=\pi{\lambda\hspace*{-0.23cm}-}^2
{\cal P}_{cap}^{\ell}(E,\ell;\alpha_1,\alpha_2)
 \end{equation}

Here ${\lambda\hspace*{-0.23cm}-}$ is the de Broglie wavelength of the 
entrance  channel.  The capture probability ${\cal P}_{cap}^{\ell}(E,\ell)$ 
 is   equal to 1 or 0 for the given beam energy and orbital angular momentum.
  Our calculations showed that in dependence on the beam energy, $E=E_{c.m.}$,
  there is a window for capture as a function of orbital angular momentum:
  \[{\cal P}^{\ell}_{cap}(E) = \left\{ \begin{array}{ll} 1,  \hspace*{0.2 cm} 
   \rm{if}
\ \  \ell_{min}<\ell<\ell_d \ \
\rm{and} \ \ {\it E>V}_{Coul}
  \\ 0, \hspace*{0.2cm} \rm {if}\ \ \ell>\ell_d  \ \
  or \ \   \ell<\ell_{min} \  \rm {and} \ \
  {\it E>V}_{Coul}
  \\ 0,  \hspace*{0.2cm} \rm{for \  all} \ \ell \ \
     \hspace*{0.2cm} \rm{if} \ \
  {\it E\leq V}_{Coul}\:,
 \end{array}
 \right.
 \]
where $\ell_{min}\ne0$ can be observed  when the beam energy
is  large  than the Coulomb barrier ($V_{Coul}$).
It means that the friction coefficient is not so strong to trap the projectile
  into the potential well (see fig.2 in ref.\cite{Nasirov}).

The number of the partial waves giving a contribution to the
capture is calculated by the solution of the equations for  the
radial and orbital motions simultaneously \cite{GiardEPJ8}. They
are defined by the size of the potential well of nucleus-nucleus
 potential $V(R,Z_1,Z_2; \{\beta^{(k)}_i\}, \{\alpha_k\})$ and the values
of the radial $\gamma_R$ and tangential $\gamma_t$ friction
coefficients, as well as by  the moment of inertia for the
relative motion. Here $Z_k, \beta^{(k)}_i$ and $\alpha_k$ are the
charge numbers, deformation parameters and orientation angles of
the symmetry axes of nuclei, $k=1,2$ and $i=2,3$ correspond to
quadrupole and octupole deformations, respectively. The
nucleus-nucleus interaction potential, radial and tangential
friction coefficients and inertia coefficients  are calculated in
the framework of our model \cite{Nasirov,Giardina,GiardEPJ8}.

\subsection{Complete fusion}
\label{fusion}

For the capture events we calculate the competition between
quasifission and complete fusion with a statistical approach
\cite{FazioMPL20}.  The competition between the complete fusion
and quasifission is obtained
 as the branching ratio between the transition  of the dinuclear
system from its position in the valley of the PES  to the ``fusion
lake" overcoming the intrinsic fusion  barrier $B^{*}_{fus}$  on
the charge number axis (complete fusion),  and the  decay  to the
``quasifission sea" after overcoming  the quasifission barrier
$B_{qf}$ on the radial distance axis.  The size of the potential
well decreases by increasing the orbital angular momentum $\ell$,
{\it i.e.} the valley of PES becomes shallow and as  result the
lifetime of the DNS decreases. At the same time the intrinsic
fusion barrier $B^{*}_{fus}$ increases while the quasifission
barrier $B_{qf}$ decreases. So, we conclude that the contribution
of the quasifission  increases by increasing  the angular momentum
for a given beam energy \cite{Giardina,FazioPRC72}.

 The intrinsic fusion barrier $B^{*}_{fus}$ for a given
projectile-target pair is the height of the saddle-point in the
valley of the PES along the axis of the DNS charge asymmetry.
 The  PES of the dinuclear
system is calculated as a sum of binding energies of interacting
nuclei,  nucleus-nucleus interaction potential $V(R)$ and
rotational energy,

\begin{eqnarray}
\label{PES}
U(Z,A,R,\{\beta^{(k)}_i\},\{\alpha_k\})&=&Q_{gg}-V^{CN}_{rot}(\ell)
+V(R,Z,Z_{CN}-Z; \{\beta^{(k)}_i\},\{\alpha_k\})\nonumber\\
&+&V^{(DNS)}_{rot}(\ell,\{\beta^{(k)}_i\},\{\alpha_k\})),
\end{eqnarray}
where  $Q_{gg}=B_1(Z_1)+B_2(Z_{CN}-Z)-B_{CN}(Z_{CN})$,
 $B_1$, $B_2$, and $B_{CN}$  are binding energies of the constituent nuclei
of DNS and compound nucleus, respectively; $Z_{CN}$ is charge number of 
compound nucleus, $V(R,Z,Z_{CN}-Z)$ is the nucleus-nucleus
 interaction potential of the DNS  nuclei;
  $V^{(DNS)}_{rot}(\ell)$ and $V^{CN}_{rot}(\ell))$
 are rotational energies of DNS and the compound nucleus.
 ${\beta^{(1,2)}_i}$ and $\alpha_{1,2}$
 are deformation parameters and orientation angles of axial symmetry axis 
 of interacting  nuclei. The binding energy values are obtained from
the tables \cite{Audi,MollNix}). The dependence of PES on the
shell structure of the nuclei forming the dinuclear system and on
the orbital angular momentum leads to the strong influence of the
entrance channel (\cite{Nasirov,Giardina,GiardEPJ8,FazioMPL20}).

The effects connected with the entrance channel appear in the
partial fusion cross section $\sigma_{\ell}^{fus}(E)$, which is
defined by the product of the partial capture cross section and
the related fusion factor $P_{CN}$ presenting the competition
between complete fusion and quasifission processes:
\begin{eqnarray}\label{sigmafus}
  \sigma^{\ell}_{fus}(E,{\alpha_i})&=&\sigma^{\ell}_{cap}(E)
  P_{CN}(E,\ell,{\alpha_i}),\\
 P_{CN}(E,\ell,{\alpha_i})&=&\sum_{Z=2}^{Z_{CN}/2}Y_Z(E)
 P^{(Z)}_{CN}(E,\ell,,{\alpha_i}),
\end{eqnarray}
where $\sigma^{\ell}_{cap}(E)$ is the partial capture cross
section which is defined by formula (\ref{parcap}). The details of
the calculation method are described in ref.\cite{FazioMPL20}.
 $P_{CN}(E,\ell)$ is  the hindrance factor for formation of compound
 nucleus connected with the  competition between
complete fusion and quasifission as possible channels of evolution
of the DNS. Due to nucleon transfer between the DNS constituents
the wide range of charge asymmetry can be populated in dependence
on landscape of the potential energy surface.
 This phenomenon is taken into account
by using of the branching ratio $P_{CN}(E,\ell)$ as the sum of
ratios of the widths related to the overflowing over the
quasifission barrier $B_{qf}(Z)$ at a given mass asymmetry,
over the inner fusion barrier $B_{fus}(Z)$ on mass asymmetry
axis to complete fusion and over $B_{sym}(Z)$ in opposite
direction
 to the symmetric configuration of DNS:
\begin{equation}
\label{pcne}
P_{CN}(E,\ell;\{\alpha_i\})=\sum_{Z=Z_{sym}}^{Z_{max}}
Y_Z(E^*_{Z})P^{(Z)}_{CN}(E^*_{Z},\ell;\{\alpha_i\})
\end{equation}
where $E^*_Z=E-V(Z,R_m,\ell;\{\beta_i\},\{\alpha_k\})+\Delta
Q_{gg}(Z)$ is the excitation energy of DNS  for a given value of
its charge-asymmetry configuration $(Z, Z_{CN}-Z)$ and
$Z_{CN}=Z_1+Z_2$;  $V(Z,R_m,\ell;\{\beta_i\},\{\alpha_k\})$ is the
minimum value of the nucleus-nucleus  potential well; the position
of the minimum  is marked as  $R=R_m$ (see fig. \ref{figDNS}c);
$\Delta Q_{gg}(Z)$ is the change of the $Q_{gg}$-value by changing
the DNS charge asymmetry; $Y_Z(E^*_{Z})$ is the probability
of population of the configuration $(Z, Z_{CN}-Z)$ at
$E^*_{Z}$, $\ell$ and given orientation angles
$(\alpha_P,\alpha_T)$. $Y_Z(E^*_{Z})$  was obtained by
solving the master equation for the evolution of the dinuclear
system (charge) mass asymmetry (for details see refs.
\cite{Nasirov,SymNas}.
 $Z_{sym}=(Z_1+Z_2)/2$~and ~$Z_{max}$~correspond
to the point where the driving potential reaches its maximum
 (see fig. \ref{figDNS}b) \cite{FazioPRC72,FazioMPL20}.

 The branching ratio $P^{(Z)}_{CN}(E^*_{Z})$ is calculated
 as a ratio of widths related to the overflowing over the
 quasifission barrier $B_{qf}(Z)$ at a given mass asymmetry,
 over the intrinsic barrier $B_{fus}(Z)$ on mass asymmetry axis
 to complete fusion and over  $B_{sym}(Z)$ in opposite direction
 to the symmetric configuration of DNS:
\begin{equation}
 \label{Gammas} P^{(Z)}_{CN}\approx
 \frac{\Gamma_{fus}(Z)}{\Gamma_{qf}(Z)+ \Gamma_{fus}(Z)+\Gamma_{sym}(Z)}.
 \end{equation}
Here, the complete fusion process is considered as the evolution
of the DNS along the mass asymmetry axis overcoming
$B_{fus}(Z)$ (a saddle point between $Z=0$ and $Z=Z_P$) and
ending in  the region around $Z=0$ or $Z=Z_{CN}$ (fig.
\ref{figDNS}b). The evolution of the DNS in the direction of the
symmetric configuration increases the number of events leading to
quasifission of more symmetric masses. This kind of channels are
taken into account by the term $\Gamma_{sym}(Z)$. One of the similar ways
was used in ref.\cite{AdamPRC69} in calculations of the evaporation
residues cross sections in the reactions with actinides.

 The widths of these "decays" leading to  quasifission and complete fusion
can be presented by the formula of the  width of  usual fission
\cite{Siemens}:
\begin{equation}
\label{Gammai} \Gamma_{i}(Z)=\frac{\rho_i(E^*_{Z})T_Z}{2\pi\rho(E^*_{Z})}
\left(1-\exp\frac{(B_i(Z)-E^*_{Z})}{T_Z}\right),
\end{equation}
where
 $\rho_i(E^*_{Z})=\rho(E^*_{Z}-B_i(Z))$;
 $B_i=B_{fus}$, $B_{qf}$, and
$B_{sym}$. $T_Z$ is the temperature of the dinuclear
system consisting of fragments with charge numbers $Z$ and
$Z_{CN}-Z$: $T_Z=\sqrt{8 E^*_{Z}/A_{CN}}$. Usually the
value of the factor
$$\left(1-\exp\left[(B_i(Z)-E^*_{Z})/T_Z\right]\right)$$ in
 (\ref{Gammai}) is approximately equal to the unit. Inserting eq. 
 (\ref{Gammai}) in
(\ref{Gammas}), we obtain  the expression (\ref{Pcn}) used in our
calculations:
\begin{equation}
 \label{Pcn} P^{(Z)}_{CN}(E^*_{Z})=\frac{\rho_{fus}(E^*_{Z})
}{\rho_{fus}(E^*_{Z}) +
\rho_{qfiss}(E^*_{Z})+\rho_{sym}(E^*_{Z})}.
 \end{equation}

\section{Results and discussion}
\label{discuss}

 In this section we compare the capture, fusion and quasifission excitation
functions with the   available experimental data which were
obtained by the analysis of the angular distribution of fission
fragments. The calculated partial capture, fusion and quasifission
excitation functions in this work are used to determine the mean
square values of angular momentum $<\ell \,^2 >$ for the dinuclear
system and compound nucleus, and to find the anisotropy $A$ of the
angular distribution by formula (\ref{anisA}). The results for $A$
are compared with the experimental data.

\begin{figure}[h,t,b]
\vspace{0.50cm}
\begin{center}
\resizebox{0.75\textwidth}{!}{\includegraphics{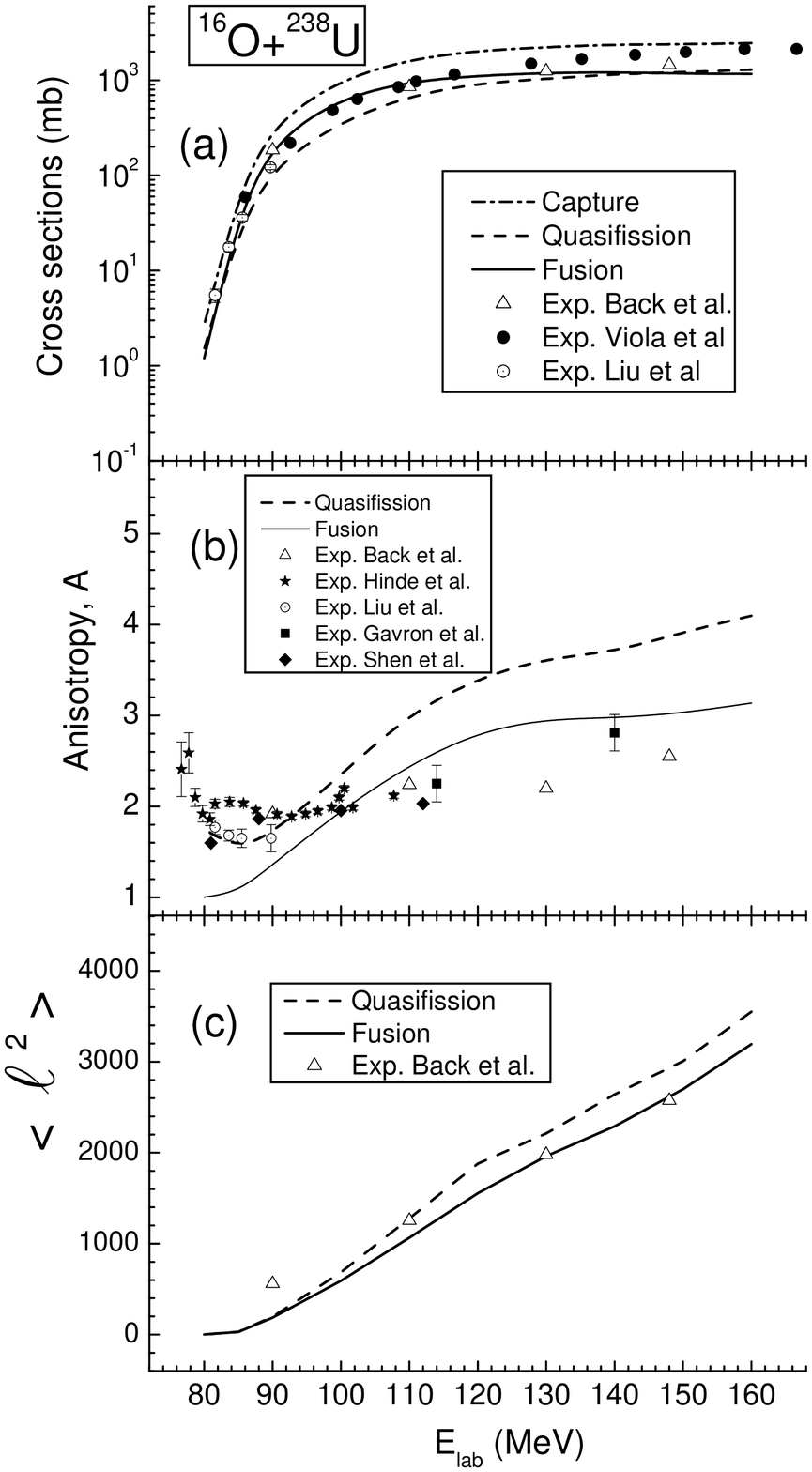}}
\vspace{-0.7cm} \caption{\label{3figO16U} (a) The calculated
capture, fusion and quasifission excitation functions for the
$^{16}$O+$^{238}$U reaction are compared with the measured fission
excitation function of refs. \cite{Back,Viola}; (b) the anisotropy
of the angular distribution obtained in this work by using of
partial fusion and quasifission excitation functions is compared
with  the experimental data from refs.
\cite{Back,Hinde53,Zhang49,Gavron}; (c) the calculated values of
$<\ell^2>$ for the $^{16}$O+$^{238}$U reaction obtained separately
for  complete fusion and quasifission in comparison with
experimental data \cite{Back}.}
\end{center}
\end{figure}

 Fig. \ref{3figO16U}a shows that at low energies the
calculated cross sections are nearly equal and are in good
agreement with the experimental data for the $^{16}$O+$^{238}$U
reaction. The measured data
 is related by a mixture of the complete fusion and quasifission
 fragments in equal proportions.
 In the energy interval  $90 <E_{lab}<130$ MeV the contribution of 
 fragments of the
 fusion-fission process  dominates over the one of quasifission
 (see fig. \ref{3figO16U}a).
 An increase of the quasifission contribution
by increasing the beam energy is explained by an increase of
events with the formation of the dinuclear system with the large
orbital angular momenta because the intrinsic fusion barrier
$B^*_{fus}$ increases and the quasifission barrier decreases  by
increasing of $\ell$ (see refs. \cite{Nasirov,FazioPRC72}).

 The authors of ref. \cite{Hinde53}  in detail analyzed the angular
 anisotropy of fragments at low energies to show the dominant role of
 the quasifission in collisions of
 the projectile with the target-nucleus  when the axial symmetry axis of
 the last is oriented along or near the beam direction. It was obtained
 large values of the anisotropy at low energies and these data
 were assumed to be connected with the quasifission because a mononucleus 
 or  dinuclear system formed in the near tip collisions has the elongated
 shape. This shape can be far from the one corresponding to the
 saddle point \cite{Hinde53}. Therefore, for such system,
  there is a hindrance at its transformation into  compound nucleus.

    The comparison of our calculated
 anisotropy values for the quasifission and complete fusion fragments
 with the ones which were presented in  refs.
 \cite{Hinde53,Back,Viola,Zhang49,Gavron} shows that the
 anisotropy $A$ connected with  quasifission  is  very close
 to the experimental data  (fig. \ref{3figO16U}b) at low energies
 $E_{lab} < 100$ MeV.   In spite of the quasifission and fusion-fission
 cross section are close, the fusion-fission fragments show less anisotropy 
 due to small $<\ell^2>$ values and large $\mathcal{J}_{eff}$.  The large 
 anisotropy for quasifission fragments
 is explained by small temperature $T_{DNS}$ and small effective moment
 of inertia $\mathcal {J}_{DNS}$.   Because the fragments under discussion
  were  formed in collisions  with  orientation angles  
  $\alpha_T \le 30^{\circ}$  for the target symmetry axis.

\begin{figure}[h,t,b]
\begin{center}
\resizebox{0.75\textwidth}{!}{\includegraphics{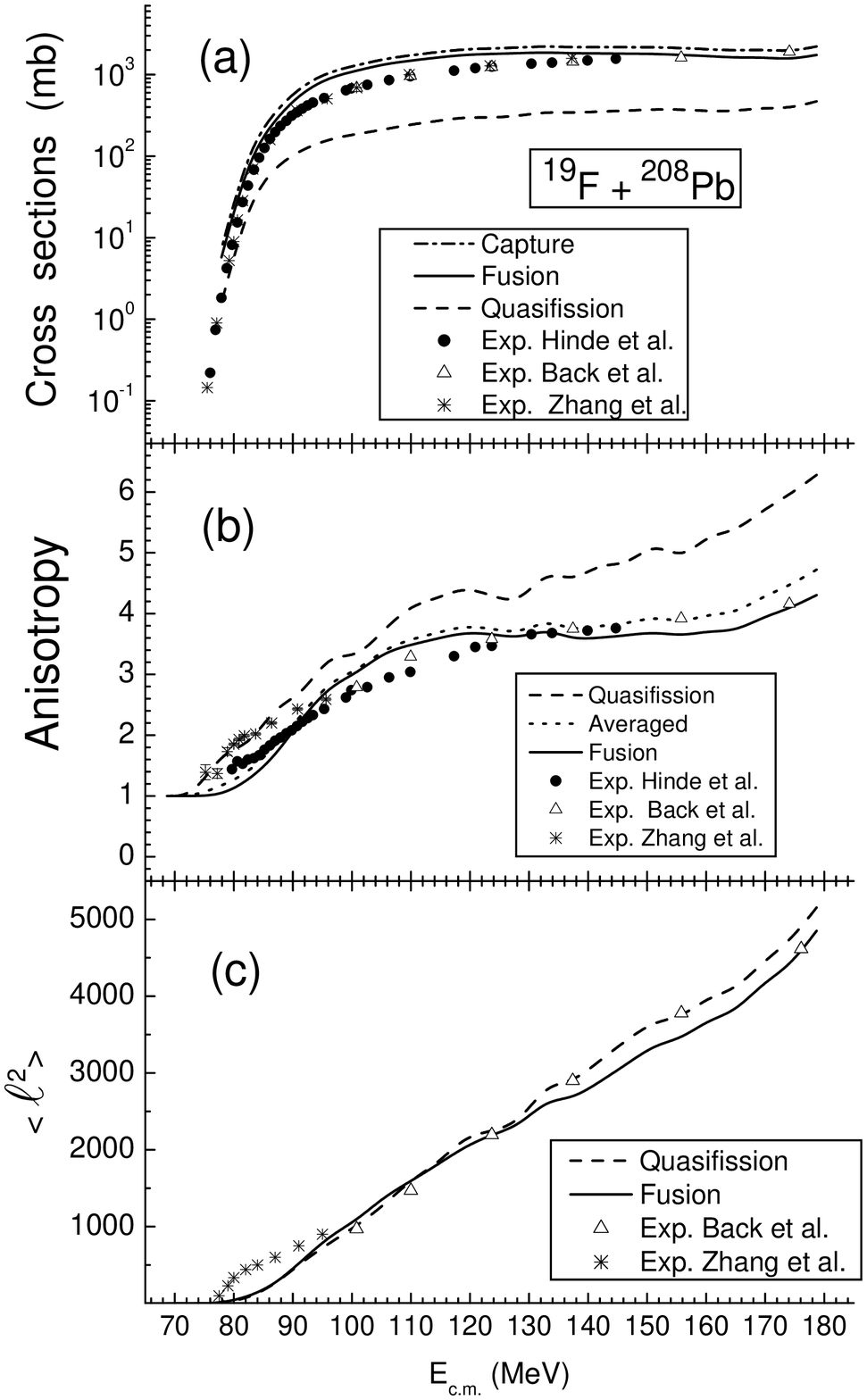}}
\vspace{-1.0cm} \caption{\label{3figF19Pb}(a) The calculated
capture fusion and quasifission excitation functions for the
$^{19}$F+$^{208}$Pb reaction is compared with the measured fission
excitation function from the  refs. \cite{Hinde,Back,Zhang} ; (b)
the anisotropy of angular distribution obtained in this work by
using of partial fusion and quasifission excitation functions is
compared with
 the experimental data from refs. \cite{Hinde,Back,Zhang} ;
  (c) comparison of the  values of $<\ell^2>$ for
 the $^{19}$F+$^{208}$Pb reaction calculated separately for the
 complete fusion and quasifission with the experimental data  refs.
 \cite{Back,Zhang}.}
\end{center}
\end{figure}

  Figure \ref{3figO16U}c shows the comparison of the calculated mean
 square of the angular momentum $<\ell\,^2>$ of the fissioning systems
 (dinuclear system  and compound nucleus) with the data extracted
 from the measured angular  distribution  of fragments in ref.\cite{Back}.
 The agreement of the fusion and quasifission angular momentum
 distributions with the experimental data is well  for all values
 of  beam energy excluding of point $E_{lab}$=90 MeV.
  The authors of
 ref. \cite{NishioPRL}  concluded that  in the sub-barrier
 region, in this reaction, the contribution of the quasifission
 is negligible. This conclusion  has been made by a comparison
 of calculated excitation functions for the  evaporation residues
 with the experimental data \cite{NishioPRL}.
 They did not need to include a hindrance to fusion
 to reproduce the experimental data.
 Authors used  the coupled-channel code CCDEGEN \cite{Hagino1},
 which is based on a version of the CCFULL code described in
 \cite{Hagino2} to calculate the fusion excitation function and
 the results were used as input for the statistical model calculation
 of evaporation residue cross sections by using the code HIVAP
 \cite{HIVAP}.

 One can see that even in the case of using a  deformed
target-nucleus, at sub-barrier energies the yield of
fusion-fission fragments are comparable  with
 the yield of quasifission fragments.
\begin{figure}[h,t,b]
\vspace{3.5cm}
\begin{center}
\hspace{-0.7cm}
\resizebox{0.95\textwidth}{!}{\includegraphics{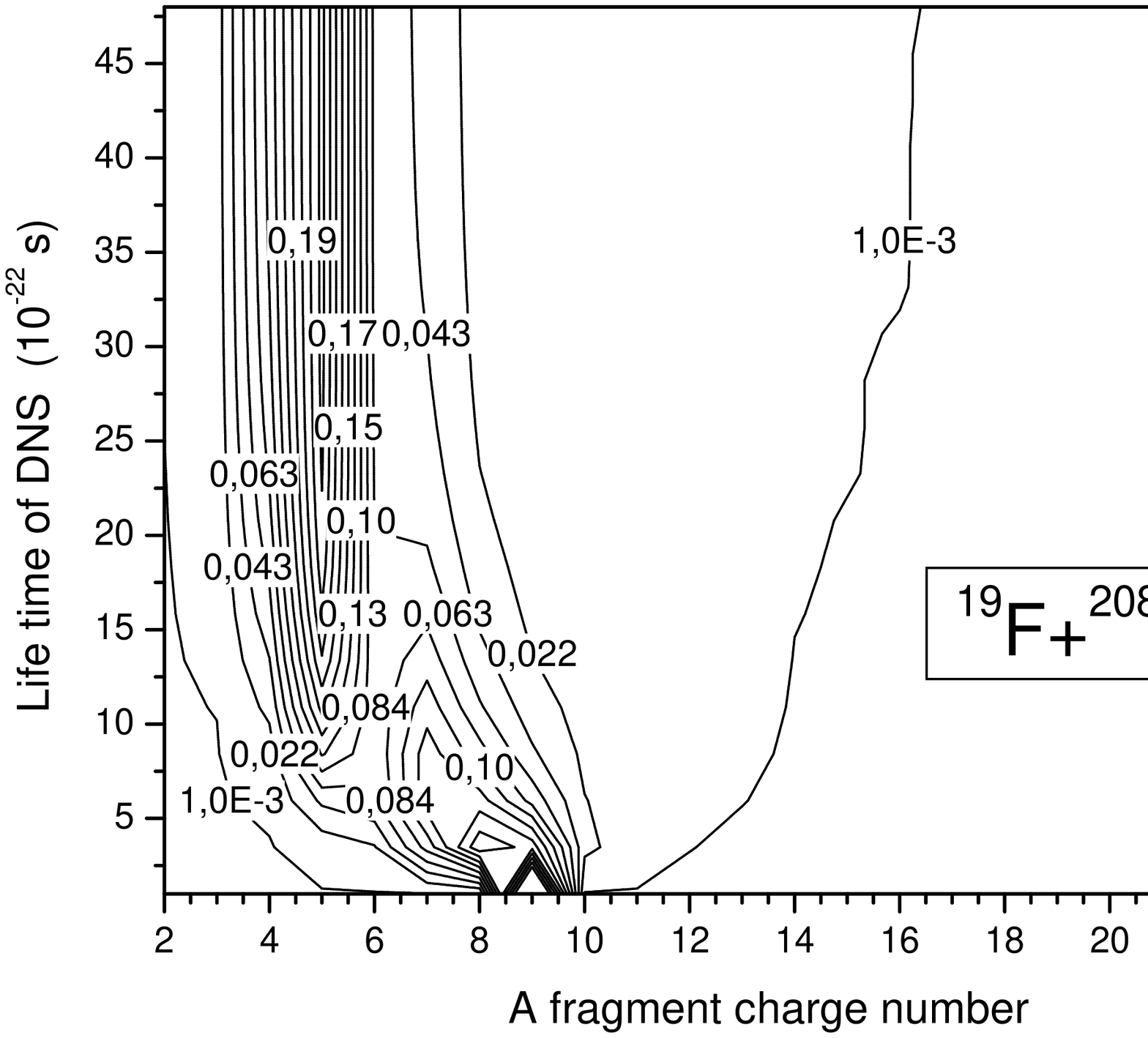}}
\vspace{-5.0cm} \caption{\label{massF19} The time dependence of
the mass distribution of quasifission and deep-inelastic transfer
processes for the $^{19}$F+$^{208}$Pb reaction.} 
\end{center}
\end{figure}

In fig.\ref{3figF19Pb}a, we show  the comparison of the excitation
functions for  the capture, complete fusion and quasifission
calculated for  the  $^{19}$F+$^{208}$Pb reaction in the framework
of our model with the experimental data of the fission cross
sections presented in refs.\cite{Hinde,Back,Zhang}. Our results
for the complete fusion are in agreement with the experimental
data. It means that
 the contribution of  quasifission in the measured
 data is small.  We should note that the maximum of  the calculated
 mass  (charge) distribution of quasifission fragments is
  near  masses of the projectile-like
 and target-like fragments of  the $^{19}$F+$^{208}$Pb reaction.
 The absent of the large anisotropy at the low energies is
 explained by the fact that the massive target nucleus has the 
 spherical shape  and ${\mathcal J}^{DNS}_{eff}$ is not so small 
 as in the   $^{16}$O+$^{238}$U reaction.

 In fig. \ref{massF19} we show the
time dependence of the charge distribution $Y_Z(t)$ for yield of
products of the quasifission processes. It was obtained by solving
of the master equation for the evolution of the dinuclear system
(charge) mass asymmetry (for details see ref. \cite{Nasirov}). It
is seen that the light fragments of quasifission have a charge
less than $Z=9$ and the asymptotic value is $Z=5$. Heaviest
fragments have a charge larger than $Z=82$ and the largest  value
is $Z=86$. This result is caused by the influence of the shell
structure of interacting nuclei. Moreover,  the fragments around
the initial charge number $Z=9$ at time $t=(5$--10)$\cdot
10^{-22}$s can be considered as fragments of deep-inelastic
collisions when the dinuclear system is formed for the short times
(no capture).

  From the good agreement of our results on the excitation function
 of the complete fusion and the angular anisotropy
 $A$ for the $^{19}$F+$^{208}$Pb reaction  (see fig. \ref{3figF19Pb}a
 and fig. \ref{3figF19Pb}b) with the experimental data presented as 
 the cross section of the fusion-fission process we can conclude that in
this reaction the fusion-fission mechanism  dominates over
quasifission mechanism.

 In fig. \ref{3figF19Pb}c  our theoretical results are compared with 
 the values of $< \ell\,^2 >$ extracted from the description of the
experimental results on the angular anisotropy $A$ of the
$^{19}$F+$^{208}$Pb reaction  \cite{Back,Zhang}. The good
agreement between the calculated and experimental results confirms
the correctness of the angular momentum distribution for the
complete fusion and quasifission calculated with our model.

The dominant role of  quasifission can be seen in the
$^{32}$S+$^{208}$Pb reaction which is a more symmetric than  the
above discussed two reactions. A sufficient role of the
quasifission in this reaction was suggested by the authors of the
experiment in ref.\cite{Back}. But they did not present
quantitative results of the ratio between complete fusion and
quasifission contributions. It is well known that this is very
complicated task due to the strong overlap in mass and angular
distributions of the fragments from both processes.

\begin{figure}[h,t,b]
\begin{center}
\resizebox{0.75\textwidth}{!}{\includegraphics{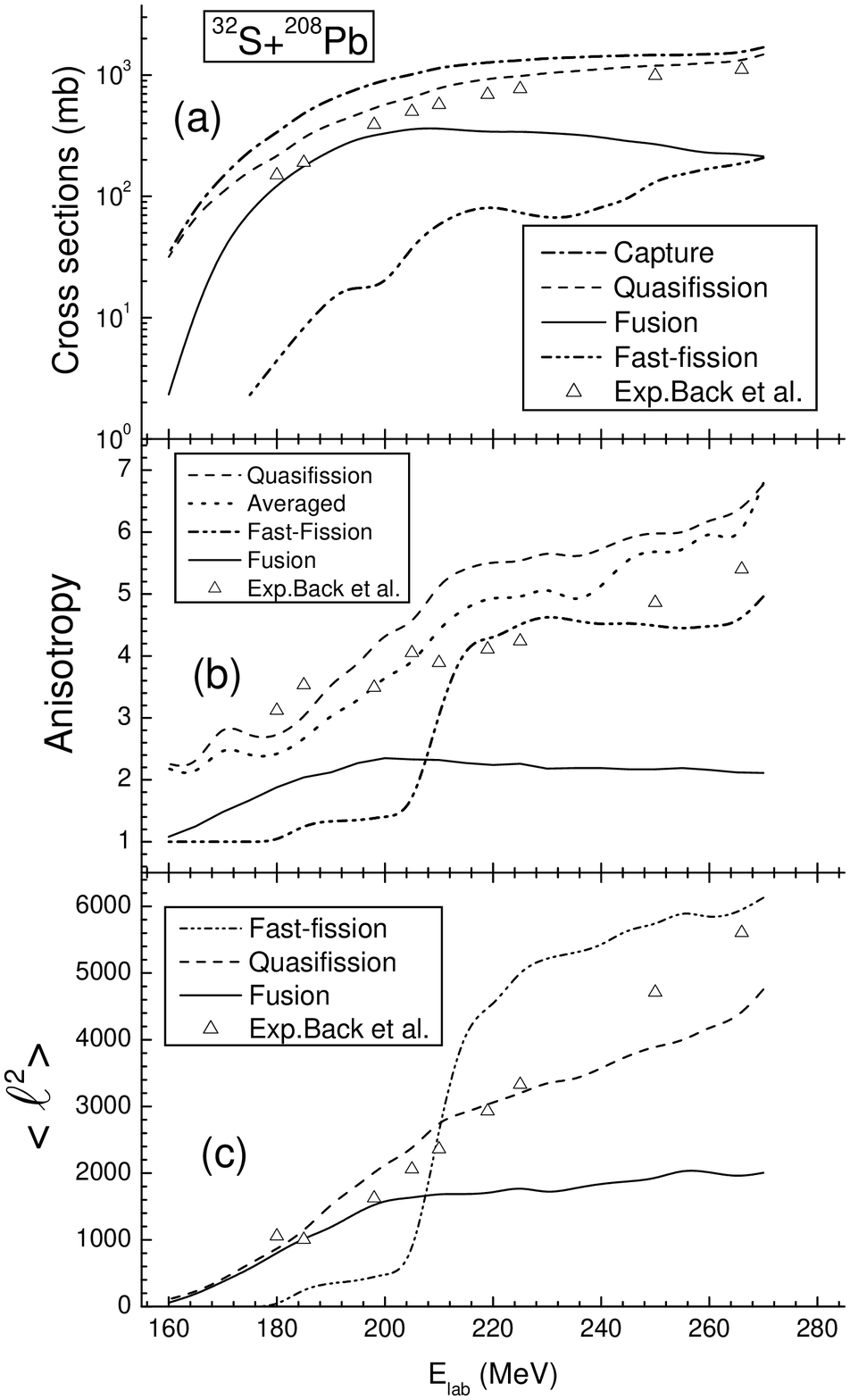}}
\vspace{-1.0cm} \caption{\label{3figS32Pb}  (a) The calculated
 capture, fusion and quasifission excitation functions for
the $^{32}$S+$^{208}$Pb reaction are compared with the measured
fission excitation function of the  ref. \cite{Back}. (b) The
anisotropy of angular distribution obtained in this work by using
of partial fusion and quasifission excitation functions is
compared with  the experimental data of ref. \cite{Back}. (c) The
values of $<\ell\,^2>$ for
 the $^{32}$S+$^{208}$Pb reaction calculated separately for the
 complete fusion and quasifission in comparison
 with the experimental data of  ref. \cite{Back}}.
\end{center}
\end{figure}

We have theoretically  analyzed the contributions of the above
mentioned processes.  In fig. \ref{3figS32Pb}a, we compare the
excitation functions for capture, complete fusion and quasifission
calculated for this reaction with the experimental data for the
fission cross section presented in ref. \cite{Back}. Our results
for complete fusion are lower than the experimental fission cross
sections. Our statement is that the data contain a large amount of
contributions of the quasifission fragments together with
fusion-fission fragments. The ratio of the yields of quasifission
fragments to fusion-fission fragments is larger at the lower and
higher beam energies.
At the low energies the competition between complete fusion and
quasifission is very sensitive to the peculiarities of  the
potential energy surface. The height of the intrinsic fusion
barrier $B^*_{fus}$ is comparable with the excitation energy of
the dinuclear system and, therefore, there is a hindrance to
complete fusion.

At the highest values of beam energy the hindrance to complete
fusion appears due to the increase of $B^*_{fus}$ as a function of
the orbital angular momentum $\ell_{DNS}$  of  the dinuclear
system. At the same time the quasifission barrier $B_{qf}$
decreases by increasing of $\ell_{DNS}$. The combined effect from
the behaviour of the $B^*_{fus}$ and $B_{qf}$ barriers as
functions of $\ell_{DNS}$ makes quasifission as the dominant
process \cite{Giardina,FazioPRC72}.
 This phenomenon is  common for all reactions with
projectiles heavier than $^{16}$O.  The fast-fission mechanism
also contributes to the anisotropy of the angular distributions of
fragments in all of the above mentioned reactions at large values
of $\ell$. According its definition the fast-fission mechanism
takes place when complete fusion occurs at large orbital angular
momentum but there is not a fission barrier for the being formed
compound nucleus. The range of angular momentum leading to the
fast-fission is $\ell_{B_{fiss}=0}<\ell<\ell_{fus}$ where
$\ell_{B_{fiss}}$ is a value at which the fission barrier for
compound nucleus disappear;
 $\ell_{fus}$ is a maximum value  of $\ell$ at which complete fusion
 takes place. The value of $\ell_{fus}$ is different for different
 orientation angles of the colliding nuclei.

 \begin{figure}[h,t,b]
\vspace{1.6cm} \hspace{-1.0cm}
\begin{center}
\resizebox{0.95\textwidth}{!}{\includegraphics{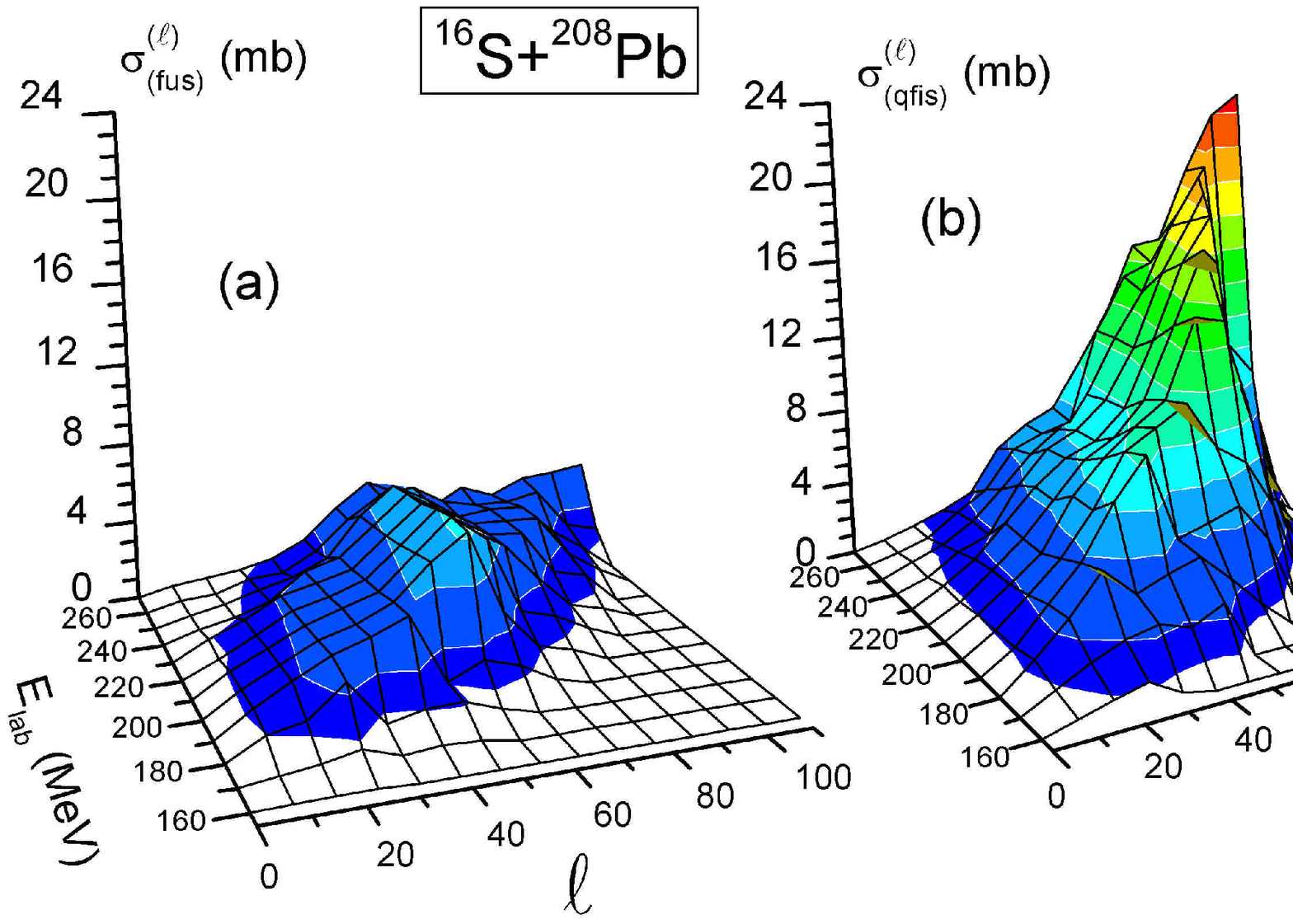}}
\vspace{-5.5 cm} \caption{\label{SpinSPb}  Angular momentum
distribution of the partial cross sections for (a) complete fusion
 and (b) quasifission calculated for the $^{32}$S+$^{208}$Pb reaction.}
\end{center}
\end{figure}

Note that quasifission can take place at small values of angular
momentum including $\ell =0$ in contrast to the fast-fission which
occurs at $\ell_B \le \ell \le \ell_{cap}$ where $\ell_B$ is the
minimum value of angular momentum of the compound nucleus where
the fission barrier disappears and $\ell_{cap}$ is the maximum
value of orbital angular momentum leading to capture. In
fig.\ref{SpinSPb}, we present as an example our results of the
angular momentum distribution of the partial fusion and
quasifission cross sections for the $^{32}$S+$^{208}$Pb reaction.

We estimated the contribution of the fast-fission in the
$^{32}$S+$^{208}$Pb reaction. As fig. \ref{3figS32Pb}a shows, the
fast-fission cross section is small in comparison with the one of
the other processes. At large beam energies, the fast-fission and
complete fusion cross sections are comparable. The
energy-dependences of anisotropy of the angular distribution of
reaction fragments are shown in fig. \ref{3figS32Pb}b.  The
contribution of fusion-fission process is small in comparison with
the data from ref. \cite{Back}.  This fact shows the dominance of
quasifission fragments in  the measured anisotropy of the angular
distribution.
 The  dotted line in
fig.\ref{3figS32Pb}b is obtained by averaging  the anisotropies
$A_{qfiss}$, $A_{fus}$ and $A_{ffiss}$ corresponding to the
quasifission, fusion-fission and fast-fission fragments,
respectively, according to the formula:
\begin{equation}
<A>=\frac{(\sigma_{qfiss}A_{qfiss}+\sigma_{fus}A_{fus}+
\sigma_{ffiss}A_{ffiss})}{(\sigma_{qfiss}+\sigma_{fus}+\sigma_{ffiss})},
\end{equation}
where $\sigma_{qfiss}$, $\sigma_{fus}$,   and   $\sigma_{ffiss}$
are the quasifission, fusion,  and fast-fission cross sections,
respectively; $A_{qfiss}$, $A_{fus}$, and $A_{ffiss}$ are the
corresponding anisotropies.

In fig.\ref{3figS32Pb}c, we compare our theoretical results with
the values of $< \ell\,^2 >$ extracted from the description of the
experimental data on the angular anisotropy $A$ of the
$^{32}$S+$^{208}$Pb reaction in ref. \cite{Back}. The extracted
values of  $< \ell\,^2 >$ from the measured data  are in good
agreement with our  theoretical results for the fusion-fission
process.  Note that at lower and higher values of the beam energy,
the curve for the quasifission is closer to the experimental data.
This confirms our conclusion of the role of quasifission made
above in the  discussion of fig.\ref{3figS32Pb}a.

\begin{figure}[h,t,b]
\vspace{2.5cm}
\begin{center}
\hspace*{-0.35cm}\resizebox{0.8\textwidth}{!}{\includegraphics{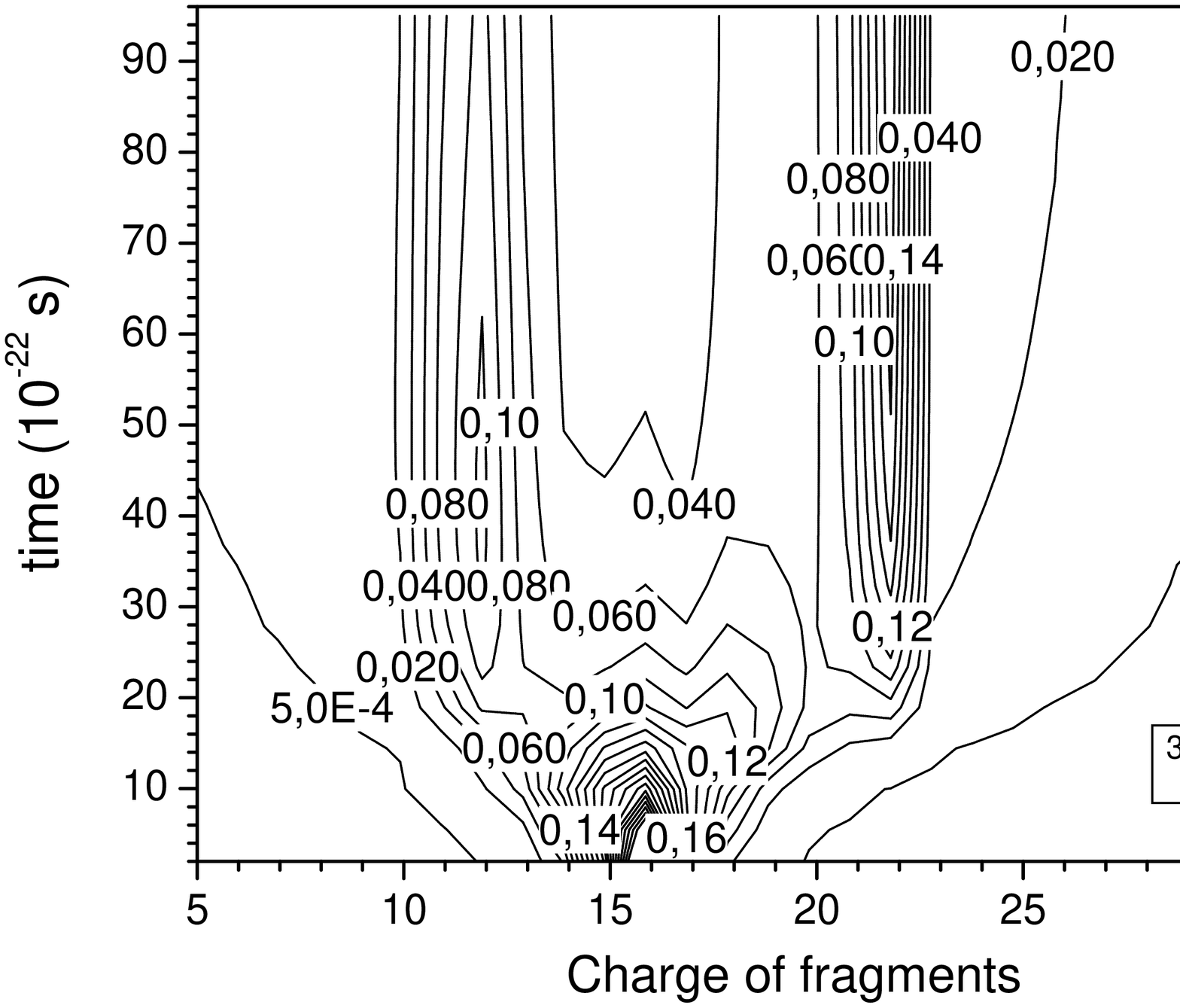}}
\vspace{-4.5cm} \caption{\label{chardisSPB} The time dependence of
the charge distribution of quasifission and deep-inelastic
transfer processes for the $^{32}$S+$^{208}$Pb reaction.} 
\end{center}
\end{figure}

  In order to comment the overestimation
of the measured data by our results for the quasifission
excitation function, we calculated of  evolution of the charge
distribution in the dinuclear system as in the case of the
$^{19}$F+$^{208}$Pb reaction. The maximum of the charge
distribution of the $^{32}$S+$^{208}$Pb reaction splits into two
peaks with increasing  interaction time. It is seen from
fig.\ref{chardisSPB} that the first maximum of the lightest
components is concentrated around the charge number $Z=12$ and the
other maximum around the charge value $Z=22$. This effect is
connected with  the influence of shell structure of the
interacting nuclei. It seems to us that the quasifission fragments
around $Z=12$ were not registered as fission fragments. This is
seen from fig. 6 of ref. \cite{Tsang}. But such fragments are
included in our results of the quasifission contribution. As a
result we overestimated the experimental data of
refs.\cite{Back,Tsang}) for the yield of binary fragments of the
full momentum transfer reactions.

 The fragments around the initial charge number $Z=16$ at time 
 $t=(5$--10)$\cdot 10^{-22}$s in fig.\ref{chardisSPB} can be 
 considered as fragments of deep-inelastic collisions when the 
 dinuclear system is formed for short time (no capture).

\section{Conclusions}
\label{concl}

The model based on the dinuclear system
\cite{Nasirov,Giardina,Volkov} was  improved to study the
influence of the quasifission on the angular anisotropy of the
measured fission-like fragments.
 We compared our calculated results with the  experimental
data on the  excitation function of the fusion-fission fragments,
anisotropy $A$ of the angular distributions and mean square values
$< \ell\,^2 >$ extracted from the description of the measured
anisotropy $A$ for the $^{16}$O + $^{238}$U
\cite{Hinde53,Back,Viola,Zhang49,Gavron}, $^{19}$F + $^{208}$Pb
\cite{Hinde,Back,Zhang} and $^{32}$S + $^{208}$Pb \cite{Back}
reactions.
 The experimental studies of the
angular distributions of fragments in heavy-ion reactions show
distinct deviation  from the SSM theory.  We explain this
deviation by  the presence of a contribution of the quasifission
reactions.

 The importance of the quasifission mechanism at the low  beam energies
was used to explain the large anisotropy  of in the angular
distribution of fragments of the full momentum transfer reaction
was discussed in ref. \cite{Hinde}. Our results obtained taking
into account contributions of different orientation angles of the
symmetry axis of the deformed $^{238}$U target to the measured
anisotropy $A$ confirm this interpretation. At the low beam
energies we observe capture (formation of the relatively long
living DNS) only for $\alpha_T \le 30^{\circ}$ and the angular
distribution of the quasifission fragments shows large anisotropy
$A$=1.7--2.0. The small values of $\mathcal{J}_{eff}^{DNS}$ and
DNS  temperature $T_{DNS}$ are responsible for this phenomenon. In
the reactions with the spherical $^{208}$Pb target,
$\mathcal{J}_{eff}^{DNS}$ is not so small to cause in  $^{19}$F +
$^{208}$Pb the similar effect observed in the $^{16}$O + $^{238}$U
reaction.
  The calculated fusion and quasifission cross sections
 are nearly equal and are in good agreement with the experimental data
 for the $^{16}$O+$^{238}$U reaction. Therefore, we conclude that
 the measured data is related by a mixture of the complete fusion and 
 quasifission  fragments.   Considering the quasifission as a
``fission" of the dinuclear system from a not compact shape we
estimate the mean square values of the angular momentum  $\ell$
and anisotropy $A$ of the angular distribution of the reaction
fragments. The experimental data of the anisotropy $A$ are
described  if we  also take into account  the contribution of the
quasifission fragments.

 For the  $^{19}$F + $^{208}$Pb reaction \cite{Hinde,Back,Zhang}
the contribution of the quasifission fragments is comparable
at low energies with the one of fusion-fission mechanism and
the last mechanism become dominant for the beam energy $E_{lab}> 90$ MeV.
So in the  $^{19}$F + $^{208}$Pb reaction  the fusion-fission 
fragments give the main contribution to the measured data.
 The effect of quasifission appears  only at more higher beam energies.

The analysis of the measured data for the  $^{32}$S + $^{208}$Pb
reaction showed the dominant role of quasifission in this
reaction. It was determined by the comparison of the calculated
fusion and quasifission cross sections, anisotropies $A_{fus}$ and
$A_{qfiss}$ connected with these processes, as well as
$<{\ell}^2>$ with the corresponding experimental data.  We
conclude that the appearance of the competition between
quasifission and complete fusion depends  on such parameters of
entrance channel of reactions as mass asymmetry, orbital angular
momentum and beam energy.

  This conclusion supports the statement  of B. Back
{\it et al.} \cite{Back}  and M. Tsang {\it et al.} \cite{Tsang}
that the assumption of  the fusion (and formation of a truly
equilibrated compound nucleus) during the first step of the
reaction is not valid in the analysis of the experimental data of
fission fragments. Therefore, these authors hypothesized the
quasifission contribution on the experimental data in order to
describe  the angular anisotropy of the detected fragments. The
good agreement of  our results with the experimental data shows
that our model can be applied to analyze the anisotropy of angular
distribution of the reaction fragments and the contribution of
quasifission fragments in the measured data which depend on the
charge asymmetry of reaction in the entrance channel,
peculiarities of the shape and shell structure of colliding
nuclei.

\section*{Acknowledgments}
 The authors are grateful to Profs. R.V. Jolos,
 and A. Sobiczewski for helpful discussions.
  This work was performed partially under the financial support of 
  the DFG, RFBR
 (Grant No. 04-02-17376) and INTAS (Grant 03-01-6417).
  The authors (A.K.N. and R.K.U.) thank DAAD and Polish-JINR cooperation 
  Program for  support while staying at the Giessen University
  and Soltan Institute for Nuclear Studies, respectively.
 A.K.N. and R.K.U.  are also grateful to the Fondazione Bonino-Pulejo
 (FBP) of Messina for the support received in the collaboration with
 the Messina group.
 \,Authors (A.I.M,  A.K.N. and   R.K.U) are  grateful to the 
 Center on Science
 and Technologies at the Ministry Cabinet (Grant No. F-2.1.8) and
 the Support Fond of Fundamental Research of the Academy of Science
 of Uzbekistan (No. 64-04) for partial support.  A.K. Nasirov
 is grateful to the Universit\'e Libre de Bruxelles for the support
 received in the collaboration.
\begin{figure}[h,t,b]
\vspace{1.8cm}
\begin{center}
\resizebox{0.75\textwidth}{!}{\includegraphics{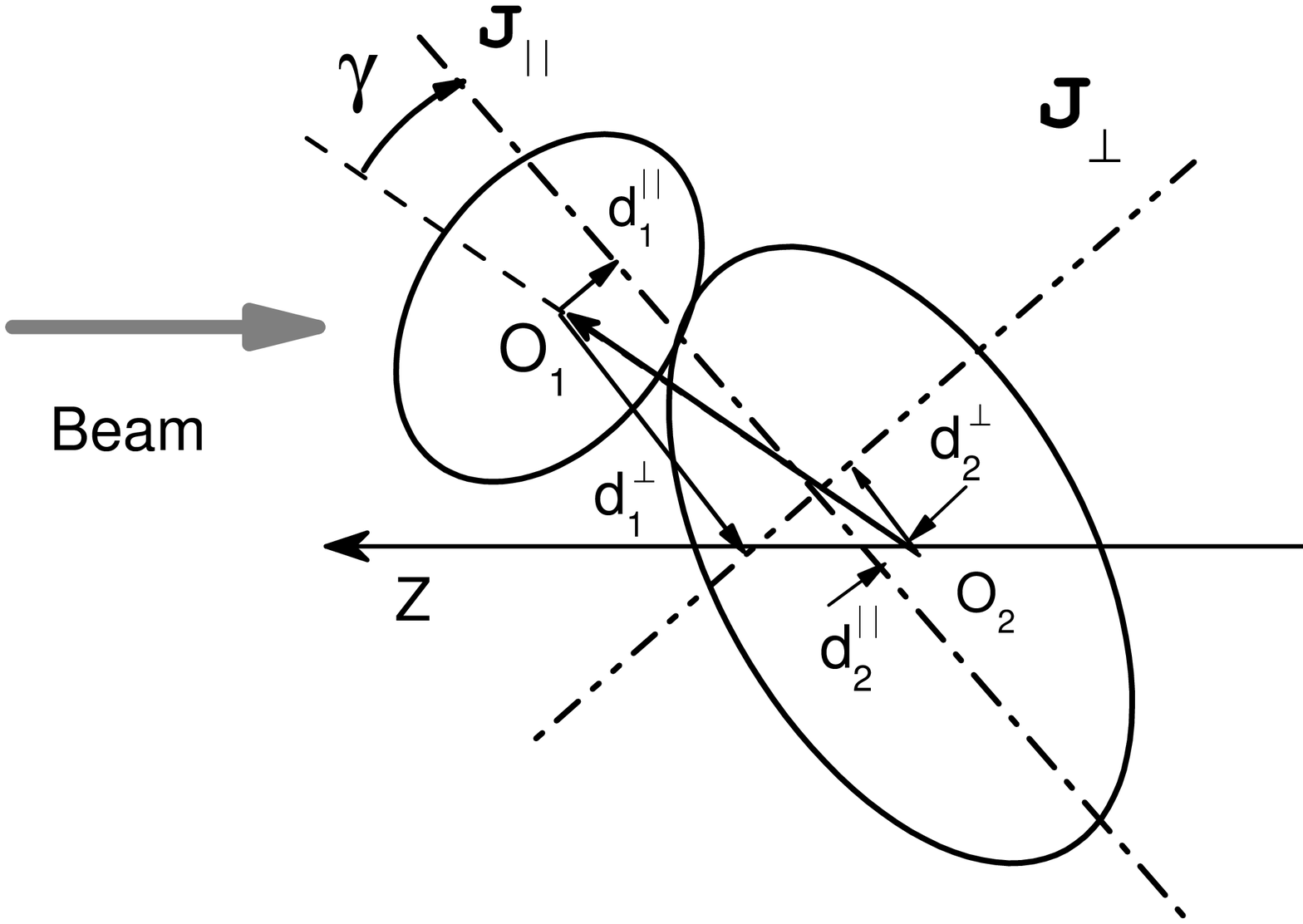}}
\vspace{-3.7cm}
\caption{\label{Jdns} The axes used to calculate the effective
moment of inertia of DNS.} 
\end{center}
\end{figure}

\appendix
\section{Calculation of the effective moment of inertia 
$\mathcal{J}^{DNS}_{eff}$  of DNS.}
 \label{ap}
The effective moment of inertia $\mathcal{J}^{DNS}_{eff}$  of DNS
which is formed in collisions with the different orientation
angles of their symmetry axes relative to the beam direction is
calculated by the formula (\ref{Jeff}).  But the moments of
inertia $\mathcal{J}^{DNS}_{\|}$ and $\mathcal{J}^{DNS}_{\bot}$
 should be the smallest and largest components, respectively.
The axis for which the value of $\mathcal{J}_{\|}$ is minimal is
found from  the condition $\frac{\partial
\mathcal{J}_{\|}}{\partial \gamma}=0$ ($\gamma$  is the angle
between the axis $\mathcal{J}_{\|}$ and beam direction)
(fig.\ref{Jdns}). The $\mathcal{J}_{\bot}$ axis is directed as a
normal to the reaction plane. The moments of inertia of nuclei are
calculated as for a rigid-body system.  For a quadrupole deformed
nucleus the moment of inertia is calculated by the expression:

\begin{eqnarray}\label{app}
 \hspace{4.2 cm} \mathcal{J}_i=\frac{M_i}{5}(a^2_i+c^2_i)
 \hspace{0.3cm} \mbox{i=1,2} 
  \nonumber \hspace{6.25 cm}({\rm A}.1)
\end{eqnarray}
where $a_i$ and $c_i$ are the  small and larger semi-axes,
respectively, of the DNS constituents  ($i=1,2$).

The moments of inertia
$\mathcal{J}^{DNS}_{\|}$ and $\mathcal{J}^{DNS}_{\bot}$ are found
by using the Steiner's theorem for the rigid-body
moments of inertia of  the DNS constituents:
\begin{eqnarray}\label{jdns1}
 \hspace{3.6 cm} \mathcal{J}^{DNS}_{\|}=\mathcal{J}_1+\mathcal{J}_2+
  M_1 d^{(1)2}_{\|}+M_2 d^{(2)2}_{\|} , \nonumber 
  \hspace{4.45 cm}({\rm A}.2)
\end{eqnarray}
\begin{eqnarray}\label{jdns2}
  \hspace{3.6 cm} \mathcal{J}^{DNS}_{\bot}=\mathcal{J}_1+\mathcal{J}_2+
  M_1 d^{(1)2}_{\bot}+M_2 d^{(2)2}_{\bot} \nonumber 
  \hspace{4.55 cm}({\rm A}.3)
\end{eqnarray}
 where $d^{(i)}_{\bot}$  ($d^{(i)}_{\|}$) is the distance between the
 center of mass of the fragment $i$ ($i=1,2$) and the axis
 corresponding to the largest (smallest)  moment of inertia of the
  dinuclear system.



\begin{thebibliography}{}
\bibitem{Ogan}
Yu. Ts. Oganessian {\it et al.}, Phys. Rev. C {\bf 70},  064609
(2004).

\bibitem{Itkis} 
M.G. Itkis {\it et al.}, Nucl. Phys. A {\bf 734}, 136147 (2004).

\bibitem{FazioEPJA22} 
G. Fazio, G. Giardina, A. Lamberto, A.I. Muminov, A.K. Nasirov, F.
Hanappe, and L. Stuttg\'e, Eur. Phys. J. A {\bf 22}, 75 (2004).

\bibitem{Hinde}
D.J. Hinde,  A. C. Berriman, M. Dasgupta, J. R. Leigh, J. C. Mein,
C. R. Morton, and J. O. Newton, Phys. Rev. C {\bf 60}, 054602
(1999).

\bibitem{Hinde53}
D. J. Hinde, M. Dasgupta, J. R. Leigh, J. C. Mein, C. R. Morton,
J. O. Newton, and H. Timmers, Phys. Rev. C {\bf 53} 1290 (1996).

\bibitem{BackPRL}
B.B. Back, {\it et al.}, Phys. Rev. Lett. {\bf 50}, 818 (1983).

\bibitem{Back}
B.B. Back, {\it et al.}, Phys. Rev. C {\bf 32}, 195 (1985)

\bibitem{Shen}
W.Q. Shen {\it et al.}, Phys. Rev. C {\bf 36}, 115 (1987).

\bibitem{Vanden}
R.Vandenbosch and J.R. Huizenga, Nuclear Fission (Academic, New
York, 1973)

\bibitem{Ramam85} 
V.S. Ramamurtby and S.S. Kapoor Phys. Rev. Lett. {\bf 54}, 178
(1985);  Phys. Rev. C {\bf 32}, 2182 (1985).

\bibitem{Vopkap95} 
D. Vopkapi\'c and B. Ivani\~sevi\'c,
  Phys. Rev. C {\bf 52}, 1980 (1995).

\bibitem{Toke}
J. T\~oke, {\it et al.}, Nucl. Phys. A {\bf 440}, 327 (1985).

\bibitem{Zhang} 
H. Zhang, Z. Liu, J. Xu, K. Xu, J. Lu  and M. Ruan,  Nucl. Phys A
{\bf 512}, 531 (1990).

\bibitem{Zhang2}
H. Zhang, Z. Liu {\it et.al}, J. of Nucl. and Radioch. Sciences
{\bf 3}, 99 (2002).

\bibitem{SierkPRC33} 
A.J. Sierk, Phys. Rev. C {\bf 33}, 2039 (1986).

\bibitem{Halpern}
I. Halpern and V.M. Strutinski, {\it Proceedings of the Second
United Nations International Conference on the Peaceful Uses of
Atomic Energy, Geneva, 1958} (United Nations, Geneva, 1958), Vol.
15, p.408.

\bibitem{Griffin}
J.J. Griffin, Phys. Rev. {\bf 116}, 107 (1959).

\bibitem{Nasirov} 
A.K. Nasirov {\it et al.}, Nucl. Phys. A {\bf 759}, 342 (2005).

\bibitem{Murakami} 
T. Murakami, C.-C. Sahm, R. Vandenbosch, D.D. Leach, A. Ray, M.J.
Murphy, Phys. Rev. C {\bf 34}, 1353  (1986).

\bibitem{Giardina}
G. Fazio {\it et al.}, Eur. Phys. J. A {\bf 19},  89  (2004).

\bibitem{FazioPRC72} 
 G. Fazio {\it et al.}, Phys. Rev. C {\bf 72},  064614 (2005).

\bibitem{GiardEPJ8} 
G. Giardina {\it et al.}, Eur. Phys. J.  A {\bf 8}, 205 (2000).

\bibitem{SymNas}
A.K. Nasirov,  G. Giardina, A.I. Muminov, W. Scheid, U.T.
Yakhshiev,  {\it  Proc. of the Symposium on Nuclear Clusters: from
Light Exotic to Superheavy Nuclei, Rauischholzhausen, Germany, 5-9
August, 2002}, ed. R. Jolos and W. Scheid, EP Systema, Debrecen,
Hungary, 2003, p.415-426; Acta Physica Hungarica A{\bf 19}  (2004)
pp.109-120.

\bibitem{Volkov}
V.V. Volkov, {\it Contributed Papers of Nucleus-Nucleus Collison II,
Visby, 1985}, edited by B. Jakobson and K. Aleclett (North-Holland,
Amsterdam, 1985), Vol. 1, p.54; Izv. Akad. Nauk SSSR Ser. Fiz.
{\bf 50}, 1879 (1986); {\it Proceedings of the International 
School-Seminar on Heavy Ion Physics, Dubna, 1986}, 
D7-87-68 (Dubna, 1987), p.528;
{\it Proceedings of the 6th International Conference on Nuclear
Reaction Mechanisms, Varenna, 1991}, edited by E. Gadioli, (Ricerca
Scientifica ed Educazione Permanente, Supplemento n.84, 1991),
p.39.

\bibitem{FazioMPL20} 
G. Fazio et al., Mod. Phys. Lett. A {\bf 20}, 391 (2005).

\bibitem{Audi} 
G. Audi, A.H. Wapstra, Nucl. Phys. A {\bf 595}, 409 (1995).

\bibitem{MollNix} 
P. Moller, J. R. Nix, W. D. Myers and W. J. Swiatecki,
 At. Data Nucl. Data Tables {\bf 59},  185
(1995).
\bibitem{AdamPRC69} 
G.G. Adamian, N.V. Antonenko, and W. Scheid,
Phys. Rev. C {\bf 69},  044601 (2004).
\bibitem{Siemens} 
 P.~J. Siemens and A.~S. Jensen,
 {\it Elements of Nuclei, Lecture Notes and Supplements in Physics},
 (Addison-Wesley, Redwood City, California, 1987).

\bibitem{Viola}
V.E. Viola and T. Sikkeland, Phys. Rev. {\bf 128}, 767 (1962).

\bibitem{Zhang49}
Z. Liu, H. Zhang, Z. Liu, J. Xu,  Y. Qiao, X. Qian, C. Lin, Phys. 
Rev. C {\bf 54}, 761 (1996).

\bibitem{Gavron}
A. Gavron {\it et al.}, Phys. Rev. Lett. {\bf 52},  589 (1983).

\bibitem{NishioPRL}
K. Nishio, H. Ikezoe, Y. Nagame, M. Asai, K. Tsukada, S. Mitsuoka,
K. Tsuruta, K. Satou, C.J. Lin, and T. Ohsawa, Phys. Rev. Lett.
{\bf 93}, 162701 (2004).

\bibitem{Hagino1} K. Hagino (unpublished).

\bibitem{Hagino2}
K. Hagino, N. Rowley, and A.T. Kruppa,
Comput. Phys. Commun. {\bf 123}, 143 (1999).

\bibitem{HIVAP}
 W. Reisdorf and M. Sch\"adel, Z. Phys. A {\bf 343}, 47 (1992).

\bibitem{Tsang}
M.~B. Tsang, D. Ardouin, C.~K. Gelbke, W.~G. Lynch, Z.~R. Xu,
B.~B. Back, R. Betts, S. Saini, P.~A. Baisden, M.~A. McMahan,
Phys. Rev. C {\bf 28}, 747 (1983).
\end{thebibliography}
\end{document}